\documentclass[twocolumn]{svjour3}          
\smartqed  
\usepackage{graphicx}
\usepackage[square,numbers]{natbib}
\usepackage{lmodern}

\usepackage{tikz}
\usetikzlibrary{quantikz2}

\usepackage[export]{adjustbox}
\usepackage{amsfonts}
\usepackage{amsmath}
\usepackage{amssymb}
\usepackage[greek,english]{babel}
\usepackage[shortcuts]{extdash}
\usepackage{hyphenat}
\usepackage{hyperref}
\usepackage[utf8]{inputenc} 
\usepackage{microtype} 
\usepackage{qtree}
\usepackage{todonotes}
\usepackage{soul}
\setuptodonotes{inline}

\renewcommand{\citet}{\cite}

\newif\ifdraftcolors

\ifdraftcolors
    \newcommand{\rev}[1]{{\color{purple}{#1}}}
\else
    \newcommand{\rev}[1]{#1}
\fi

\journalname{Submitted manuscript awaiting}

\begin{document}

\title{Quantum Natural Language Processing}


\author{Dominic Widdows \and
        Willie Aboumrad \and
        Dohun Kim \and
        Sayonee Ray \and 
        Jonathan Mei
}


\institute{IonQ, Inc. 4505 Campus Dr, College Park, Maryland, USA \\
\email{[widdows$|$jonathan.mei$|$aboumrad$|$sayonee]@ionq.com} \\
Pohang University of Science and Technology \\
\email{dohunkim@postech.ac.kr} \\
}

\date{Preprint of paper submitted to \href{https://link.springer.com/journal/13218/updates/26098708}{Springer KI Special Issue}. Please check there for final edited version.}

\maketitle

\begin{abstract}

Language processing is at the heart of current developments in artificial intelligence,
and quantum computers are becoming available at the same time. This has led to 
great interest in quantum natural language processing, and several early proposals and experiments.

This paper surveys the state of this area, showing how NLP-related techniques have been used in
quantum language processing. We examine the art of
word embeddings and sequential models, proposing some avenues for future investigation and discussing the tradeoffs present in these directions. We also highlight some recent methods to compute attention in transformer models, 
and perform grammatical parsing. We also introduce a new quantum design for the basic task of text encoding 
(representing a string of characters in memory), which has not
been addressed in detail before.


Quantum theory has contributed toward quantifying uncertainty and explaining ``What is intelligence?''
In this context, we argue that ``hallucinations'' in modern artificial intelligence systems 
are a misunderstanding of the way facts are conceptualized:
language can express many plausible hypotheses, of which only a few become actual.


\end{abstract}

\section{Introduction}
\label{intro}

In early 2024, quantum computing and AI are two of the most rapidly-moving and talked-about
areas of science and technology. 
The availability of dialog systems based on
large language models (LLMs) has raised the profile of natural language processing (NLP)
to a historic high, developing and expanding very quickly.
This expansion has led to AI models being deployed as systems and introduced as components
in new ways, leading to improvements and efficiencies, but also mistakes and concerns.
Thus the demand for improvements in AI is at an all-time high, with a renewed focus
on reliability and trust. 

Quantum theory offers new forms of mathematical modeling, computation and communication. 
Mathematical models for language operations motivated explicitly by quantum theory have been
used in information retrieval \citep{rijsbergen2004geometry,sordoni2013modeling}, 
logic and disambiguation \citep{widdows2003wordvectors}, and language composition \citep{clark2007combining,coecke2010distributional}.
Similar models have been developed in many social sciences and demonstrated successful 
results over classical alternatives, long before any such models were implemented and run on 
quantum computers.
More abstractly, entire classes of quantum machine learning (ML) models have been theoretically shown to have more 
expressive power than comparable classical models \citep{coyle2020born,yu2023provable},
\rev{though this does not guarantee improvements in results more generally \citep{bowles2024better}}. 
Running basic NLP algorithms on quantum has become possible only in the
last few years, with early-stage results reported by \citet{lorenz2023qnlp,widdows2024near}.

This paper is intended as an introduction to this landscape, for those interested in 
language processing and quantum computing, but not necessarily specialists in either.
Firstly, Section \ref{sec:qc-intro} gives a brief introduction to quantum gates and circuits.
Section \ref{sec:qpostr} continues with an idealized example of how quantum gates could be used to represent
a text string of exponential length in a register of qubits, including some caveats and pitfalls.
This gives a glimpse of some of the wonder,
and some of the challenges, of quantum computing. 
The main body of the paper surveys ways in which other aspects of language processing have
already been modeled on quantum computers, including embedding vectors, sequences, attention, and 
grammatical structure (Sections \ref{sec:embeddings}--\ref{sec:syntax}). 
This gives a snapshot of 
of where quantum NLP has got to at this stage of the NISQ era.
Finally, Section \ref{sec:facts} discusses the challenges of choosing and distinguishing
between the hypothetical and the actual. This has taken on fresh urgency in 
AI systems for fact-checking, to avoid mistaking so-called hallucinations for assertions.
We note that language models are {\it designed} to produce both hypothetical and actual
statements, and that quantum mechanics is a better starting point than classical 
mechanics for modeling this.

\section{Quantum Computing Basics}
\label{sec:qc-intro}

In early 2024, quantum computers are real and in regular use, and quantum runtime is
offered as-a-service by many companies, via the internet / cloud. 
This section introduces some of the basic building blocks of quantum computing,
from the perspective of a developer designing quantum programs, particularly 
to run on today's noisy-intermediate scale quantum (NISQ) hardware.
The development process involves specifying a register of qubits,
and saying what logic gates and measurements should be performed on these qubits. 

The material here overlaps with the introduction of \citet{widdows2024financewalks}.
Some familiarity with quantum mechanics, especially Dirac notation, is assumed,
so that $\ket{0}$ and $\ket{1}$ are the basis states for a single qubit whose state is represented in the complex vector
space $\mathbb{C}^2$, a 2-qubit state is represented in the tensor product space $\mathbb{C}^2\otimes \mathbb{C}^2 \cong \mathbb{C}^4$
with basis states $\ket{00}, \ket{01}, \ket{10}$ and $\ket{11}$, 3-qubit states are represented in $\mathbb{C}^{\otimes 3} \cong \mathbb{C}^8$
with basis states $\ket{000}, \ket{001}, \ldots, \ket{111}$, and so on. 
For introductions to how linear algebra is written and
used in quantum mechanics, see \citet[Ch 2]{nielsen2002quantumcomputation}.
Quantum measurement is probabilistic:
if $\ket{\phi}$ is an eigenvector of a given measurement operator, then
a system in the state $\ket{\psi}$ is observed to be in the state $\ket{\phi}$ with probability given by the square of their scalar product, $\braket{\phi}{\psi}^2$ (the Born rule), and if this outcome is observed, the system is now in the state $\ket{\phi}$.

\begin{figure}
    \centering

\begin{tabular}{ccc}
Pauli-$X$ (NOT)
& 
\begin{tikzcd} \qw & \gate{X} & \qw \end{tikzcd}
\begin{tikzcd} \qw & \targ{} & \qw \end{tikzcd}
&
$\begin{bmatrix} 0 & 1 \\ 1 & 0 \end{bmatrix}$
\\[6 ex]

Hadamard ($H$)
& 
\begin{tikzcd} \qw & \gate{H} & \qw \end{tikzcd}
&
$\frac{1}{\sqrt{2}}\begin{bmatrix} 1 & 1 \\ 1 & -1 \end{bmatrix}$
\\[6 ex]

\begin{tabular}{c}
$R_X$ rotation \\
$\exp(-i\frac{\theta}{2}X)$
\end{tabular}
& 
\begin{tikzcd} \qw & \gate{R_X(\theta)} & \qw \end{tikzcd}
&
$\begin{bmatrix} \cos \frac{\theta}{2} & i\sin \frac{\theta}{2} \\ i\sin \frac{\theta}{2} & \cos \frac{\theta}{2} \end{bmatrix}$

\end{tabular}    
    \caption{Single-qubit gates used in this paper, and their corresponding matrices,  
    which operate on the superposition state $\alpha\ket{0} + \beta\ket{1}$ written as the column vector $\begin{bmatrix} \alpha & \beta \end{bmatrix}^T$.}
    \label{fig:single-qubit-gates}
\end{figure}
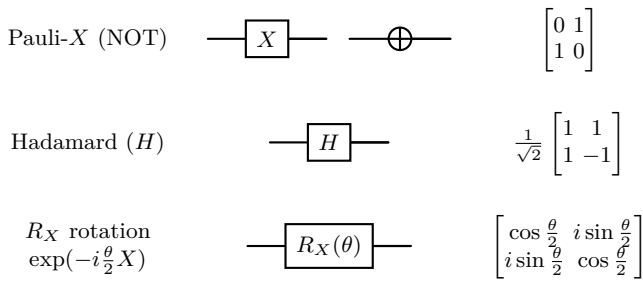

In mathematical terms, the key features that distinguish quantum from classical computers
are superposition and entanglement. 
Superposition can be realized in a single qubit: the state $\alpha\ket{0} + \beta\ket{1}$ is a superposition
of the states $\ket{0}$ and $\ket{1}$, where $\alpha$ and $\beta$ are complex numbers, with $|\alpha^2| + |\beta^2| = 1$.
Each single-qubit logic gate is a linear operator that preserves the orthogonality of the basis states and this normalization condition, 
and the group of such operators is $U(2)$, the group of complex $2\times 2$ unitary matrices. Single-qubit gates
that feature prominently in this paper are shown in Figure \ref{fig:single-qubit-gates}. So single-qubit gates coherently manipulate
the superposition state of an individual qubit.

Entanglement is a property that connects different qubits. Since the 1930's, quantum entanglement has gone from 
a hotly-disputed scientific prediction, to a statistical property demonstrated with large ensembles, to a connection
created between pairs of individual particles, to a working component in quantum computers.
All modern quantum computers have some implementation of an entangling gate, and only one kind is really needed,
because all possible 2-qubit entangled states can be constructed mathematically by combining appropriate single-qubit gates
before and after the entangling gate. Furthermore, a single 2-qubit entangling gate and a set of single-qubit gates forms 
a {\it universal gateset} for quantum computing \citep[\S4.5]{nielsen2002quantumcomputation}.
Entanglement is the crucial feature that distinguishes quantum computing algorithmically, 
because predicting the probability distributions that result from quantum operations with entanglement can become exponentially hard 
for classical computers. In simpler terms, quantum computing is special because it offers special kinds of interference, not because it 
offers special kinds of in-between-ness.

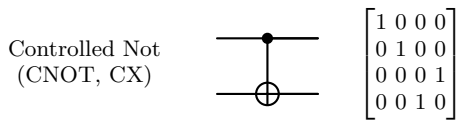
\begin{figure}
    \centering
\begin{tabular}{ccc}
\begin{tabular}{c}
Controlled Not \\ (CNOT, CX)
\end{tabular}
& 
\begin{tikzcd} 
\qw & \ctrl{1} & \qw \\
\qw & \targ{} & \qw
\end{tikzcd}
&
$\begin{bmatrix}
1 & 0 & 0 & 0 \\
0 & 1 & 0 & 0 \\
0 & 0 & 0 & 1 \\
0 & 0 & 1 & 0 
\end{bmatrix}$
\\

\end{tabular}    
    \caption{Two-qubit CNOT (controlled-$X$) and gate}
    \label{fig:cnot-gate}
\end{figure}

A quantum circuit consists of a register of qubits, and a sequence of logic gates that act on these qubits.
Some of the basic gates used in this paper are shown in Figures \ref{fig:single-qubit-gates} and \ref{fig:cnot-gate}.
The Pauli-$X$ gate is commonly used to flip a qubit between the $\ket{0}$ and $\ket{1}$ gates, which is why it is also sometimes called the quantum NOT gate. $X$-gates applied to different qubits can be used to prepare
an input state representing a binary-valued vector: the state 
$\ket{010 \cdots 001}$ is prepared by applying an $X$-gate to each of the qubits to be switched to the $\ket{1}$ state.

The Hadamard (H) gate is commonly used to put a qubit into a superposition state: for example, it maps a qubit
prepared in the state $\ket{0}$ to the superposition $\frac{1}{\sqrt{2}}(\ket{0} + \ket{1})$. 
Applying an H-gate to each qubit in an array is used to initialize a binary vector all of whose coordinates have
a 50-50 chance of being observed in the $\ket{0}$ or $\ket{1}$ state.

Other probabilities, anywhere in the range $[0, 1]$, can be arranged by using fractional rotations
(which might involve sending just the same laser-pulse instructions, but for different time periods). 
An example is given in the $R_X(\theta)$ gate. 
Several variational quantum algorithms work by gradually optimizing such $\theta$ parameters.

The CNOT gate is a 2-qubit entangling gate, that acts upon the
    state $\alpha\ket{00} + \beta\ket{01} + \gamma\ket{10} + \delta{\ket{11}}$. In the standard basis, its behavior can be described
    as ``performing a NOT operation on the target qubit if the control qubit is in state $\ket{1}$''.


\begin{figure}
    \centering
\begin{tabular}{ccc}
\begin{tabular}{c}
Multi- \\ Controlled \\ Not
\end{tabular}
& 
\begin{tikzcd} 
\qw & \ctrl{2} & \qw \\
\ghost{X} & \ctrl[open]{1} & \qw \\
\qw & \targ{} & \qw
\end{tikzcd}
=
\begin{tikzcd} 
\qw & \qw      & \ctrl{2} & \qw & \qw \\
\qw & \gate{X} & \ctrl{1} & \gate{X} & \qw \\
\qw & \qw      & \targ{}  & \qw & \qw
\end{tikzcd}
\\

\end{tabular}    
    \caption{Three-qubit multi-controlled gate (Toffoli gate) with $\ket{1}$ and $\ket{0}$ control states.}
    \label{fig:multi-cnot-gate}
\end{figure}
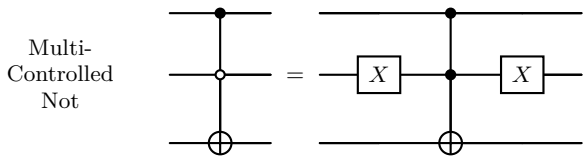

By assembling several 1- and 2-qubit gates, more qubits can become entangled, and we can define
{\it multi-controlled} gates. For example, the 3-qubit Toffoli gate in Figure \ref{fig:multi-cnot-gate}
flips the lowest qubit if the top qubit is in the $\ket{1}$ state and the middle qubit in the $\ket{0}$ state.
It is relatively easy mathematically to start arranging such gate recipes into higher-level  
operations: for example, the 3-qubit gate acts as a logical AND, from which simple binary arithmetic can be 
constructed. 

However, developers must be careful when using such constructions, because 
gate complexity and errors can easily build up. In practice, it takes 5 CNOT gates and 9 single-qubit gates
to assemble the 3-qubit Toffoli gate, so if gate errors are above 1\%, the error-rate in a 
circuit with just 3 Toffoli gates would be over 50\%. In 2024, gate error rates tend to be much better than this,
but still, typical circuits today do not reliably run more than a few hundred gates. 
NISQ-era quantum circuit development tends to tradeoff between sophistication (more gates introduces more 
tunable parameters) and reliability (fewer gates gives fewer errors).
\rev{It also leads to designs where classical components are relied upon for many parts of an NLP
pipeline, such as storing weights and comparing scores \citep{widdows2024near}.}

\rev{
In quantum machine learning,  variational circuits, or parametrized quantum circuits (PQCs), are a particularly clear example of 
such tradeoffs \cite[Ch 5]{schuld2021machine}. Variational circuit designs
use i. a quantum circuit with parameters $\{\theta_i\}$, often implemented as variable gate angles, 
that can be optimized according to a given loss function; and ii. a classical optimizer, responsible for 
evaluating the measurement outputs of the quantum circuit, and proposing updates to the parameters $\{\theta_i\}$.
When both classical and quantum components play such a prominent role, the combination is sometimes called 
a hybrid system or hybrid workflow.
}

\rev{
The transition from NISQ to fault-tolerant quantum computing will be gradual, and arguably is already underway: 
recent months have seen encouraging progress in quantum error correction and memory fidelity \citep{bravyi2024high}.
This raises long-term expectations that quantum computers will optimize crucial matrix operations,
such as solving systems of equations \citep{harrow2009quantum}, and quantum singular value transformation \citep{gilyen2019quantum}.
However, it is important to remember that even fault-tolerant quantum computing will come with serious caveats, 
and in particular, quantum components bring no advantage if their I/O costs outweigh their computational gains \citep{aaronson2015read}.

Through the rest of this paper, we highlight examples of some of these considerations and design differences as they appear.
}

\section{A Quantum String Encoding Example}
\label{sec:qpostr}

Character and string encoding is one of the most basic tasks in language processing, and
this section gives a worked example of how this might be performed on a quantum computer using some of the standard quantum circuit and gate patterns introduced in the previous
section. 
This makes a good case-study of some of the promise and challenges
of quantum computing, and (as far as we know) is the first such proposal for
representing text strings in a quantum computer, comparable to the use of ASCII or Unicode specifications in mainstream classical computing. 

\rev{To encode a text of meaningful length, this encoding would require many layers of 2-qubit gates, and
this design would require error correction, rather than being a NISQ-era proposal.}
Other methods of encoding the meanings of texts in quantum NLP work have been devised, such as vector embeddings 
for use in machine learning classifiers, and several such techniques will be surveyed in later sections.
The example quantum circuit designs in this section are 
for the (much older) protocol of representing words as a sequence of characters chosen from a relatively small character set. 
Some of the quantum word-encoding models based on a sequences and embeddings will be surveyed in later sections of this paper.

The established way to define a string in computer science is to rely on an encoding standard such as ASCII which identifies letters with numbers (A=65, B=66, etc.), and then a string such as {\it CAB} can be represented as the number sequence [67, 65, 66].
Here the quantum developer faces an immediate and typical challenge: arrays and lists 
are not standardized components, and a strategy to read from the next location in memory needs to be introduced as part of the design.
This lack generalizes: there is much more established literature, software, and hardware for quantum algorithms than for 
quantum data structures.

\begin{figure}[t]
    \centering
\begin{tabular}{ccc}
\begin{tikzcd} 
\qw & \gate{X} & \qw \\
\qw & \ghost{X} & \qw
\end{tikzcd}
&
\begin{tikzcd} 
\qw & \ghost{X}  & \qw \\
\qw & \gate{X} & \qw 
\end{tikzcd}
&
\begin{tikzcd} 
\qw & \gate{X} & \qw \\
\qw & \gate{X} & \qw 
\end{tikzcd} 
\\
Encodes $a=01$ & Encodes $b=10$ & Encodes $c=11$ \\  
\end{tabular}
    \caption{Simple encodings for a three-letter alphabet in a two-qubit register, using the convention that the top
    qubit in the register is the least-significant ``units'' bit in the binary encoding.}
    \label{fig:simple_encodings}
\end{figure}
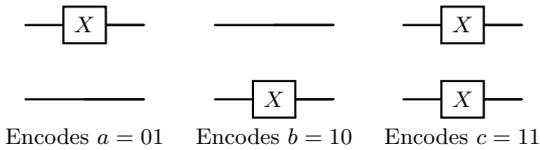

One data structure we can use as a building block in quantum circuits is binary positional notation for integers. For example, 
in a $4$-qubit register, the state $\ket{1010}$ could represent the decimal number 10 (if read left-to-right) or 5 (if read right-to-left).
This convention was used right at the beginning of quantum computing, in Feynman's proposal of how to build a quantum adder circuit \citep{feynman1985quantumcomputers}, and is part of Shor's integer factoring algorithm \citep{shor1994algorithms}.
These numbers can be represented in quantum circuits using $X$-gate bit-flip operations on the corresponding qubits, 
as in Figure \ref{fig:simple_encodings}.

Our string encoding design
works by entangling a ``position'' register that records where a character appears, along with an ``alphabet'' register that says which character appears in that position. 
So instead of a sequence $[3, 1, 2]$, the string is represented as a 
tensor product $1_P \otimes 3_C + 2_P \otimes 1_C + 3_P \otimes 2_C$, where $n_P$ is the state 
representing the $P^\mathrm{th}$ position, and $m_C$ is the state representing the $C^\mathrm{th}$ character.
A similar pattern is used by \citet{amankwah2022quantum} for representing images, and the encoding is called 
QPIXL. QPIXL uses one register to specify the location of a pixel in an image, entangled with another register saying which channel (e.g., red, green, or blue) is being referred to. So, QPIXL also keeps ``what it is'' and ``where it is'' in separate registers and entangles these. 
Emphasizing this similarity, we call our string encoding protocol {\it QPOSTR}, pronounced ``Q-poster'', 
meaning ``Quantum Positional String''.

\begin{figure}[t]
    \centering
    \includegraphics[width=\linewidth]{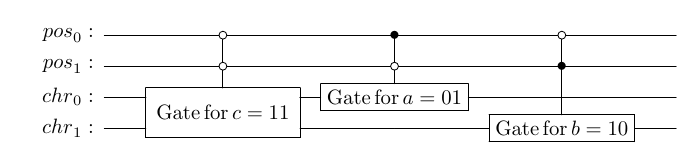}
    \caption{Position and Character Encoding for the string {\it cab}.}
    \label{fig:cab_fragment}
\end{figure}

The implementation of this formula as a quantum circuit component 
for encoding the string {\it cab} is outlined in Figure \ref{fig:cab_fragment}.
The top 2 qubits form the “position” register, and the values on the control qubits are represented by the open and closed circles, $\circ=0$, $\bullet=1$. Starting with the top as the least significant bit, 
the control states are $\circ\circ=00=0$, $\circ\bullet=01=1$, $\bullet\circ=10=2$, etc.
The bottom 2 qubits form the ``character'' or ``alphabet'' register. Each letter in the alphabet is mapped to a number corresponding to its position, so the gate-recipe for each character is like one of the simple circuits shown in Figure \ref{fig:simple_encodings}. 
To encode the string {\it cab}, a character register of 2 qubits suffices. To encode 26 letters, a character register of 5 qubits would be
required (since $2^5 = 32$), and for the ASCII character set, 7 qubits would be needed.

\begin{figure}[t]
    \centering
    \includegraphics[width=\linewidth]{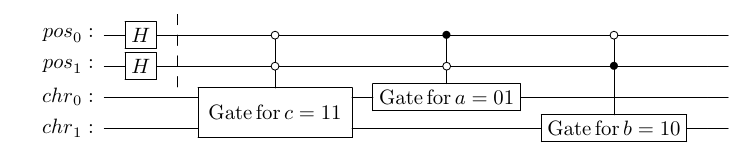}
    \caption{QPOSTR Encoding for the string {\it cab}.}
    \label{fig:cab_circuit}
\end{figure}

If the circuit above is prepared in the conventional $\ket{0000}$ state, the first gate controlled on the $\circ\circ=00$ state is the only one active, and the circuit output will be 00 (in the position register) and 11 (in the character register), saying just ``the zeroth character was a {\it c}.'' 
To prepare a superposition of the characters in all of the positions, the position register is prepared in a uniform distribution over all the available character positions, using a standard array of Hadamard (H) gates. 
This gives the full circuit for encoding the string {\it cab} in Figure~\ref{fig:cab_circuit}.
Character positions beyond the length of the string are untouched,
or left with character ``0'' in that position. Using the convention that
character ``0'' represents a space, this is equivalent to 
padding a string with trailing zeros to make its length a power of 2.


With $n$ qubits, the position register can encode up to $2^n$ positions, and with $m$ qubits, the alphabet register can encode up to $2^m$ characters. Thus, a QPOSTR circuit with $m+n$ qubits can represent a string of length up to $2^n$, with up to $2^m$ characters. By contrast, a classical computer requires $m \times 2^n$ classical bits to store the same string. 


As a thought experiment, we can use GPT-3 training metadata to demonstrate this savings. The size of the training dataset is reported at $\sim$300B tokens \citep{brown2020language}, so a generous estimate of 12 characters-per-token
allows for $12\times 300 \times 10^9 < 2^{42}$ character-positions, 
for which the positional encoding fits in 42 qubits. There are currently 149,813 Unicode characters, 
so even this alphabet fits in 18
qubits, which means that the entire $\sim$45TB training dataset of GPT-3 could fit into a mere $18+42=60$ qubits!

\begin{figure}[t]
    \centering
    \includegraphics[width=\linewidth]{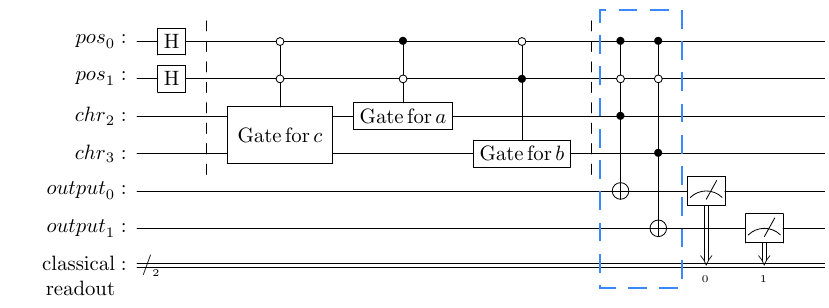}
    \caption{QPOSTR readout circuit which recovers the character {\it ``a''} from position $\circ\bullet = 01$ of the string {\it cab}.}
    \label{fig:cab_readout_circuit}
\end{figure}

We can recover information about which character is in which position by adding an output register with the same number of qubits as the character register. The circuit for this is shown in Figure \ref{fig:cab_readout_circuit}. 
The multi-controlled gates that connect the QPOSTR representation to the readout register are configured to detect the same position in the string. Each extra qubit in the controls for these gates is set to detect a particular bit in the character register, and if this bit is set to a 1, then the corresponding output bit is set to a 1.

If this process could be repeated many times, and if the output qubits could be reset to $\lvert 0 \rangle$ independently of the other qubits in the register, eventually the entire input string or any parts of it could be read out to classical memory.
Superficially, this extension of the readout circuit gives a circuit that demonstrates 
that the whole original string can be recovered from the QPOSTR quantum encoding. 
In other words, it looks as if we can recover an exponentially-long string ($2^n$ characters) 
from an encoding that uses $n + \lceil \log_2(\text{alphabet size}) \rceil$ qubits!

However, this is deceptive: due to the $H$ gates at the beginning, we have no way of knowing {\it which} position will 
be measured by each readout operation, and to guarantee statistically that each position is sampled,
we would need an exponential number of measurements on different copies of the state. (In practice, many shots of the quantum 
circuit would need to be run.)
This is in line with a general theorem in quantum computing, the Holevo bound \cite[12.1.1]{nielsen2002quantumcomputation}, 
which limits the amount of classical information that can reliably be recovered from a quantum state.


Thus, QPOSTR gives an exponential space advantage over the classical alternative.
However, it still takes a length of time at least linear in the length of the string to run the quantum circuit that loads the string into quantum memory. 
Like many quantum encodings, 
QPOSTR only gives an exponential space saving, which offers no obvious practical advantage without a corresponding time saving. Thus, even if we were to encode the GPT-3 training data, it is unclear at this moment what savings could come from doing so. 
That said, we optimistically speculate that future algorithms or data structures may be able to better take advantage of this encoding.


More generally, the challenge of preparing a quantum memory that can be maintained and successively queried is sometimes
described as research in QROM and QRAM \citep{giovannetti2008quantumram,babbush2018encoding}. 
Minimizing the number of gate operations
is a key goal in such work, and the position encoding used in QPOSTR can be regarded as one of the ``simple (but suboptimal)'' 
encoding methods described in Fig 3 of \citet{babbush2018encoding}. The task they are interested in is encoding electronic spectra, but they 
also consider encoding ``words'' as an example toy problem. The extra step that QPOSTR takes is explicitly to map register values to alphabetic 
characters, which enables such a unitary positional encoding to represent a text as a sequence of alphabetic characters.

Rather than demonstrating a new quantum advantage, the QPOSTR example is intended to showcase some of the excitement, but also some of the 
gotchas of quantum computing. It is astonishing that an exponentially long string can be encoded like this at all,
but once the engineering caveats around that statement are properly understood, we see that the explicit information we can 
recover from this representation is much smaller.

\section{Word Embeddings and Text Classification}
\label{sec:embeddings}

Representing words as vectors of coordinates is a technique that goes back at least to the 1960s and early information retrieval systems \citep{salton1983introduction}. 
The key theoretical motivation behind such
distributional semantics methods
is that words that appear in similar contexts tend to have similar meanings \cite{wittgenstein1953investigations,firth1957synopsis}. 
Based on their distribution in text, embedding techniques map words to vector spaces, where their similarity is typically measured by the inner product of their corresponding vectors. 

Semantic properties of vectors in lower-dimensional projections were analyzed in the 1990s \citep{landauer1997solution}, and by the early 2000s, overlaps between the logic of word vectors in information retrieval and state vectors in quantum mechanics had been explicitly recognized \citep{rijsbergen2004geometry,widdows2004geometry}.
In the past decade, embeddings for classical NLP have jumped from having a resurgence in academia to becoming massively mainstream in industry \citep{mikolov2013w2v, bridgwater2023rise, metinko2023pinecone}. 
Naturally, this suggests embeddings could be just as central for QNLP, especially since the mathematics of vectors and tensors
has become a common language for both AI and quantum computing \citep{widdows2021quantummathematics}.

There are many ways to add a quantum flavor to embeddings. 
In information retrieval, \cite{sordoni2013modeling}
used density matrices and quantum probability to include 
term-term dependencies in retrieval weighting. Their quantum 
probability model for bigrams prefigures the more general
probabilistic models developed by \cite{bradley2020interface}.

Word2ket \citep{panahi2019word2ket} was introduced as 
a quantum-inspired solution for compressing embeddings. 
A tensor network is a decomposition of a high-dimensional tensor
into an approximate product of lower-dimensional tensors or vectors. 
For example, if $M\approx U\otimes V$ then $M$ can be
represented using approximately $\dim(U) + \dim(V)$ coordinates, rather than $\dim(U) \times \dim(V)$.
 (More precisely, in matrix coordinates, $M'\approx uv^{T}$ for column vectors $u$ and $v$ corresponding to appropriately reshaped $U$ and $V$, and $M'$ is a reshaped version of $M$; see \cite{hitchcock1927expression,van_loan2000ubiquitous} for details.) 
Word2ket uses tensor networks to create low-dimensional approximations for individual word vectors, and entire vocabularies.
This mathematical initiative continues: for example, \cite{tomut2024compactifai} report using tensor networks
to compress the parameters of an LLM (LlaMA-2 7B model) to 30\% of its original size while retaining over 90\% of the original accuracy.

These methods are quantum-inspired, in the sense of drawing deliberately 
on quantum mathematical models, but running on classical computers.

\subsection{Building Quantum Embeddings}

Next we discuss techniques relating to embeddings intended for use on actual quantum devices, treating separately the topics of building quantum embeddings and using them. 
\rev{The circuits proposed in this section are intended for NISQ-era rather than fault-tolerant devices.}

One of the most well-known recent techniques for building word embeddings is word2vec 
\citep{mikolov2013w2v}.
Taking inspiration from this line of work, we propose a quantum computing implementation of word2vec.

Word2vec is a group of word embedding methods that use shallow neural networks to capture the semantic properties of words. Word2vec includes two popular methods: Continuous Bag-of-Words (CBOW) and Skip-gram. 
CBOW and Skip-gram have different ways of learning the word embeddings. CBOW takes the context words around the target word as input and tries to predict the target word. Skip-gram does the opposite: it takes the target word as input and tries to predict the context words. The output of both methods is a probability distribution over all of the words in the vocabulary, 
which is computed using the softmax function. This can be computationally expensive and impractical when the size of vocabulary is large.

Skip-gram with Negative Sampling (SGNS) \citep{mikolov2013distributed} is a variant of Skip-gram that reduces computational complexity by simplifying the objective function. Instead of predicting the probability over complete vocabulary, SGNS only tries to distinguish the true context words from a few randomly sampled negative words, which are assumed to be irrelevant to the target word. Thus, the classification problem is simplified from multi-class to binary. The negative sampling procedure also effectively balances the training dataset.

\begin{figure}
    \centering
    \includegraphics[width=0.7\linewidth]{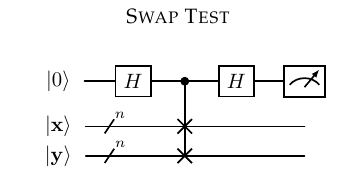}
    \caption{Swap test circuit, where the probability of measuring a \ket{0} in the top qubit is $|\braket{x}{y}|$
    reflecting the overlap between the \ket{x} and \ket{y} states \cite{barenco1997stabilisation}.}
    \label{fig:swap_test}
\end{figure}

In our implementation, we use quantum states as word vectors, and use quantum fidelity to apply cosine similarity. That is, encoding two words as $|x\rangle$ and $|y\rangle$, we measure their similarity as $|\langle x | y \rangle|^2$ via the swap test \citep{barenco1997stabilisation}, as shown in Figure \ref{fig:swap_test}.

One of the challenges of quantum word embedding is how to efficiently load words as quantum states. We consider two potential schemes for embedding words to quantum state: memory-efficient embedding and circuit-efficient embedding. 

In memory-efficient embedding, the quantum state of every word in vocabulary of size $N$ is represented by a single unitary operation ${U(\theta) \in SU(2^n)}$, where ${n=\lceil \log_2 N \rceil}$ and $\theta$ is a set of learnable parameters. The $m$-qubit quantum state for the $k$-th word ${\ket{w_k}\in \mathbb{C}^{\otimes m}}$ is obtained from $U(\theta)$ by applying it to the computational basis state $\ket{k}$ and discarding ancillary ${n-m}$ qubits. 

This scheme allows us to store a large number of words in small number of qubits, which is exponentially efficient in memory usage. However, the resulting quantum circuit that has sufficient expressiveness to implement $U(\theta)$ has exponential depth, making it impractical for circuit-based quantum computation. Moreover, the state preparation process involves post-selection, and is thus non-deterministic due to the probabilistic nature of measurement.

In contrast, the second scheme, circuit-efficient embedding, represents the $k$-th word by a quantum state of the form ${\ket{w_k}\coloneqq U(\theta_k)\ket{0} \in \mathbb{C}^{\otimes m}}$, 
where $U(\theta_k) \in SU(2^m)$ is a unitary operation parameterized by $\theta_k$, which is specific to each word. This allows us to prepare the quantum state using a depth-efficient circuit in a deterministic process, without using excessive ancillary qubits. 
While it requires more classical memory to store the parameters, it is more flexible since one can add or remove words from the vocabulary during training.

\begin{figure}[t]
    \centering
    \includegraphics[width=\linewidth]{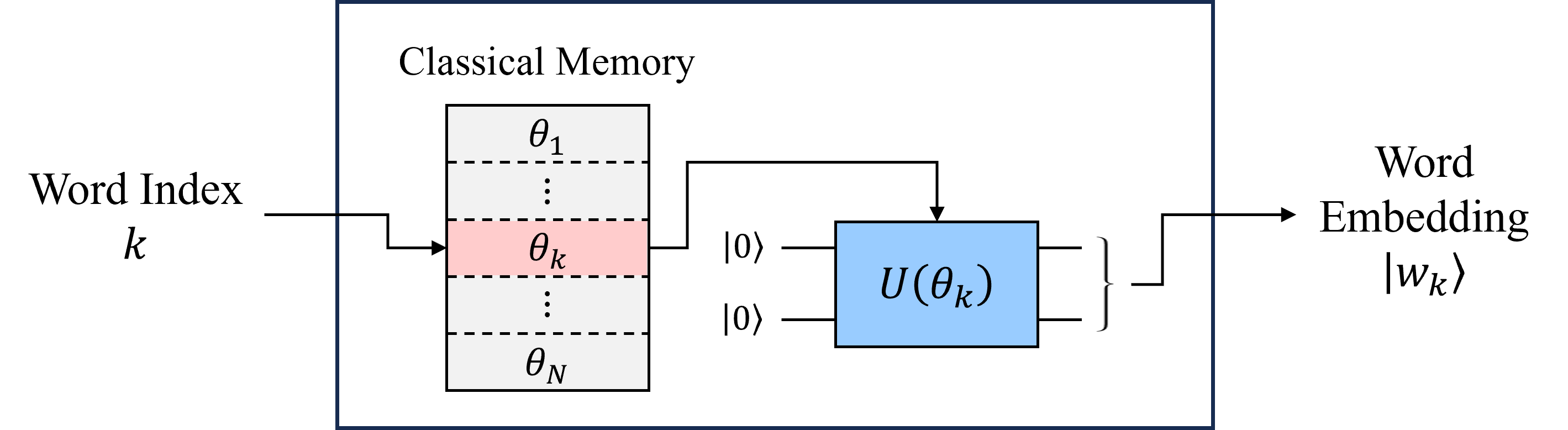}
    \caption{Circuit-efficient embedding}
    \label{fig:circuit-efficient-emb}
\end{figure}

\begin{figure}[t]
    \centering
    \includegraphics[width=\linewidth]{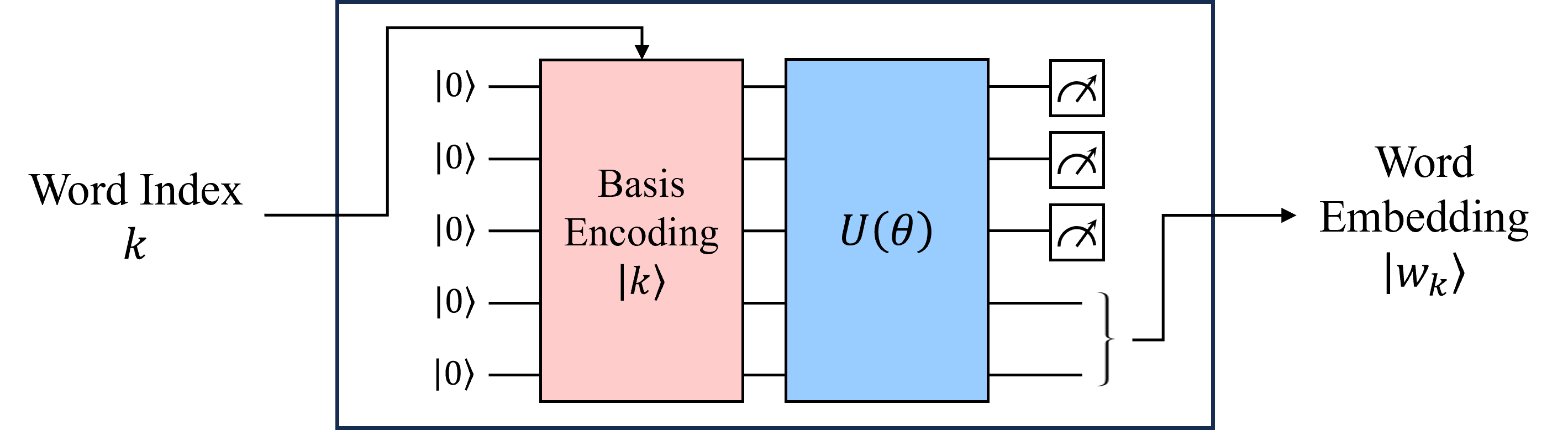}
    \caption{Memory-efficient embedding}
    \label{fig:memory-efficient-emb}
\end{figure}

The circuit-efficient and memory-efficient patterns are depicted in Figures \ref{fig:circuit-efficient-emb} and \ref{fig:memory-efficient-emb}. In structure, the circuit-efficient pattern is like the word-embedding in  
word2ket \citep{panahi2019word2ket}, and the memory-efficient pattern is like the whole vocabulary encoding. 
For word2ket, the motivation for expressing a whole vocabulary in a more entangled tensor network 
is to reduce classical memory, whereas in this design, it makes more use of the efficient quantum memory. 

To learn the parameters for these embedding schemes, we adapt the classical word2vec methods, CBOW, Skip-gram and SGNS, to the quantum setting. For quantum CBOW and Skip-gram, we introduce a parameterized unitary operator ${V(\phi)\in SU(2^n)}$ that defines the probability distribution $p_\phi(k|w) \coloneqq \lvert\bra{k}V(\phi) (\ket{w}\otimes\ket{0}) \rvert^2$, 
where $w$ is either a pooled (e.g. averaged) embedding of context words for CBOW, or the embedding for the target word for Skip-gram. This removes the need of computing the costly softmax function, by using the natural output of the quantum circuit to predict the index among $2^n$ computational basis states.
We note that quantum CBOW, cannot be directly applied for quantum word embeddings, since the direct averaging of quantum states is not a natural operation for quantum computers. Instead, it may be possible to use superpositions of quantum states.

Quantum CBOW and Skip-gram inherit the difficulties of the multi-class problem from the classical version. Both also suffer from the training problem that does not affect the classical versions: barren plateaus \citep{mcclean2018barren}. The barren plateau describes the phenomenon where the loss function and its gradient exponentially concentrate as the number of qubits increases. The unitary $V(\phi)$ leads to a barren plateau in the loss landscape, making the training process difficult to scale.

Hence, we propose quantum SGNS, which leverages a simplified structure to mitigate the scaling issue. Quantum SGNS uses the embeddings for two words directly, instead of needing to additionally train $V(\phi)$. Given the target word $\ket{w}$, quantum SGNS tries to maximize the likelihood ${p(v|w)\coloneqq \lvert \braket{v}{w} \rvert^2}$ if $v$ is a context word and minimize it if $v$ is negative sample. By combining quantum SGNS with the circuit-efficient embedding scheme, we enable their practical use on current quantum devices.
Future work includes exploring the effects of different similarity kernels on quantum word2vec, extending these circuits to implement word2ket, and understanding algebra in the embedding space.


\subsection{Using Quantum Embeddings}

Once words are encoded as vectors, these vectors can be used in many machine learning
systems, including support vector machines, which can be used for supervised classification.
This sometimes involves calculating a {\it kernel} function, which computes similarities
between input vectors, sometimes involving computations that would be intractable if all the coordinates were
constructed and compared explicitly \cite[Ch 5]{geron2019hands}.
This is regarded as a promising research direction for quantum machine learning, because quantum kernel
circuits that compare $2^n$ coordinates or amplitudes can be implemented using just $n$ qubits \citep{schuld2021machine}. 
However, as with the QPOSTR string encoding example in Section \ref{sec:qpostr}, if the number of {\it gates}
still scales with the number of coordinates, the use of a logarithmic number of {\it qubits} is a saving of space but with no corresponding time advantage.

Several quantum vector encodings or ``feature maps'' for word embedding vectors were compared by
\cite{alexander2022quantumtext}, and used for sentiment analysis experiments. The ZZ-feature map
was found to be the most successful, achieving a classification accuracy of 62\% on classification 
experiments involving small test sets of roughly 10K words each. This result showed initial promise,
and was the largest quantum text experiment reported to-date, but also indicates how small today's 
quantum NLP experiments, are compared with even modest-sized classical NLP systems.

One additional use case for embeddings is in factual grounding and retrieval, which we discuss in more detail in Section \ref{sec:facts}. 

\section{Sequential Models for Text Generation}
\label{sec:seq}

Quantum generative modeling is still a largely unexplored area of opportunity,
with many unsolved challenges.
In some cases, the data is too large and is partitioned into segments that are correlated but treated as independent for sake of computation \citep{huang2021experimental}.
In others, the term is used to describe settings in which a quantum circuit is used as a discrete source of randomness within an otherwise classical neural network that memorizes a select few data samples \citep{rudolph2022generation}.

In NLP specifically, \cite{bradley2020interface} describes how to model the joint distribution $p(X_0, X_1)$ of a given a set of bigrams $(x_0, x_1)$ and compute the marginal distributions $p(X_0)$ and $p(X_1)$ using linear algebra operations native to quantum computing (here we use capital letters to denote random variables and lower-case to denote particular values that the random variable can take). \cite{widdows2024near} follows this theory to implement a Quantum Circuit Born Machine (QCBM) \citep{benedetti2019generative} to learn the joint distribution $p(X_0, X_1)$. While the QCBM can efficiently sample pairs from $p(X_0, X_1)$ or marginals $p(X_0)$ and $p(X_1)$, generating text sequentially from this model requires sampling from the conditional $p(X_1|X_0=x_0)$. This requires discarding samples for which $X_0\ne x_0$, which can become prohibitive when scaling to larger vocabularies, and especially for rare prefix words $x_0$. In addition, the bigram model forgoes a hidden state like those found in Recurrent Neural Networks (RNNs) that can learn to represent longer dependencies. Thus, while this model can easily sample bigrams directly, it is not optimized for the task of sampling longer sequences.

A class of Bayesian network models, called n-gram models, have been successful in multiple language processing tasks, including information retrieval, text generation, and part-of-speech tagging \citep{jurafsky2023speech}. 
However, they suffer from poor performance and generalizability issues in sparse data regimes and fail to capture nonlocal syntactic relations. 
Handling out-of-vocabulary words or resolving ambiguity also pose challenges as n-grams do not have built-in 
semantic understanding \citep{bengio2000neural}. 
A Hidden Markov models (HMM) is a Markov model whose output from any given state is probabilistic rather than deterministic,
which hides the internal state. A canonical example is when the hidden states are part-of-speech tags, such N(oun) or V(erb),
which generate explicit words with given probabilities \cite[\S 9.2]{manning1999foundations}.

The feed forward and recurrent neural networks are specific instances of the HMM. However, HMMs also face challenges in estimating accurate probabilities when the context size increases or when the correlations get too long in languages. Beyond language processing, HMMs have been widely used in other scientific fields as well. In \cite{durbin1998biological}, the authors discuss probabilistic models, particularly HMMs and their derivatives, and their applications in biological sequence analysis. Language models, particularly those developed for aligning and comparing sequences, can be designed to recognize patterns in biological sequences, infer evolutionary relationships, and identify functional elements. 

Recently, quantum techniques are being explored along this direction due to their potential in capturing long-range correlations. 
\cite{gao2022enhancing} introduced a quantum enhanced version of the HMM, named the basis-enhanced Bayesian Circuit, which leverages quantum contextuality and non-locality to boost the expressivity of classical HMMs. They developed a minimal quantum extension of the bigram HMM, by incorporating measurements in a Bayesian circuit in multiple bases. They demonstrated improved performance of the quantum enhanced model in certain sequential datasets, including one containing DNA sequences with non-local structures. This leads to many interesting questions about the potential and utility of similar quantum methods in natural language processing tasks. Particularly, if quantum properties like contextuality and non-locality still give a provable advantage in terms of model expressivity when processing and extracting semantic meaning from a long sequence of words or protein structures.
 
\begin{figure*}
    \centering
    \includegraphics[width=\linewidth]{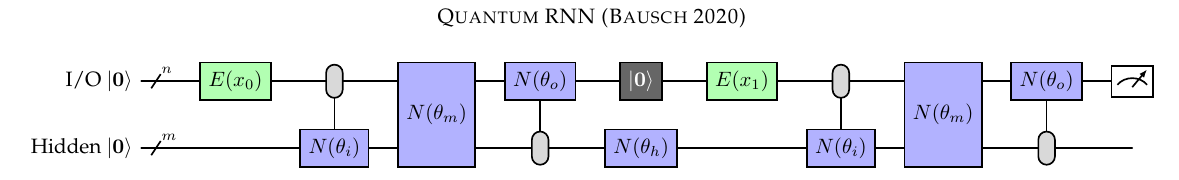}
    \caption{Bausch (2023) circuit. The input is the sequence $(x_{0}, x_{1})$. The mixing and hidden blocks are prohibitive for current hardware. The output is the next token in the sequence as the outcome from a single shot.}
    \label{fig:bausch_circuit}
\end{figure*}

\begin{figure*}[t]
    \centering
    \includegraphics[width=0.5\linewidth]{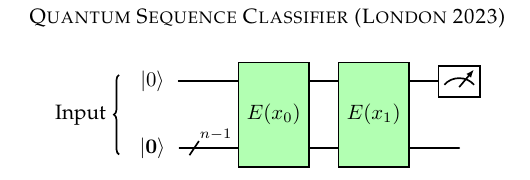}
    \caption{London et al. (2023) circuit. The input is the sequence $(x_{0}, x_{1})$. Only the first qubit is measured. The output is the probability that the sequence is in class 1.}
    \label{fig:london_circuit}
\end{figure*}

\begin{figure*}[t]
    \centering
    \includegraphics[width=0.7\linewidth]{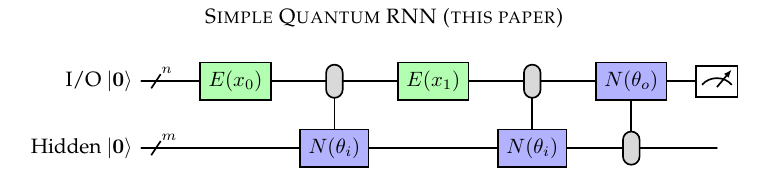}
    \caption{Proposed circuit. The input is the sequence $(x_{0}, x_{1})$. The output can be the next token in the sequence as the outcome from a single shot.}
    \label{fig:proposed_circuit}
\end{figure*}

Building toward longer sequences, \cite{karamlou2022quantum} train a quantum classifier for predicting the topic of sentences and describe a classical scheme for using the classifier to perform conditional generation (that can also be used on classical classifiers). Since it does not actually train a standalone quantum generative model, this method suffers from a similar problem of also needing to discard many samples from a base model in order to generate one sample from an induced model, though the correlated editing-based annealing scheme could be guided to be more efficient than independently sampled shots from a QCBM.
\cite{bausch2020recurrent}, shown in Figure~\ref{fig:bausch_circuit}, proposes a framework for a quantum RNN that can be used to build an actual generative model to autoregressively produce text. This autoregressive modeling allows for dealing with correlations across segments of larger data, addressing the problem from \cite{huang2021experimental}. However, the architecture, while expressive, is far too expensive for current hardware on non-trivial problem sizes. 
\cite{london2023peptide} perform sequence classification on actual hardware as shown in Figure~\ref{fig:london_circuit}. Unfortunately, the architecture they propose is not powerful enough to perform autoregressive modeling.

We explore how to bridge some of the gap between general but expensive sequence processing of \cite{bausch2020recurrent} and the currently-achievable but underpowered architecture of \cite{london2023peptide}. To achieve this, we use the paradigm of \cite{gili2023introducing} to combine the power of nonlinear Multi-Layer Perceptrons (MLPs) with the inherent randomness of quantum computing to directly sample from rich classes of probability distributions. Compared to \cite{bausch2020recurrent}, our proposed  architecture, shown in Figure \ref{fig:proposed_circuit}, drops the use of the reset operation and uses identity mixing, similar to \cite{london2023peptide}.

In the figures, the details of the quantum neurons have been abstracted away to highlight the main similarities and differences between the methods. The text labeling on the left denotes the purpose of the register. $E(x)$ denotes a block of gates for encoding inputs $x$, $N(\theta)$ denotes a nonlinear neuron parameterized by variables $\theta$ as in \cite{cao2017quantum}, and the dark gray gate with the $\ket{0}$-ket is a reset operation. The parameter subscript $i$ denotes input, $o$ denotes output, $h$ denotes hidden, and $m$ denotes mixing. The light gray vertical rounded rectangle denotes that the qubits in the register are being used as control for the corresponding neuron as a sequence of single-control gates, not as true multi-control gates. We depict the architectures from \cite{bausch2020recurrent} and \cite{london2023peptide} with a sequence length of 2 with input sequence $(x_{0}, x_{1})$, and the Multi-Layer Perceptron (MLP) from \cite{gili2023introducing} with a single hidden layer.

In simulation, the model is trained using backpropagation from gradients computed from noiseless state vector simulation. The model produces the probability distribution over the 11 words in the vocabulary corresponding to the word that the model predicts comes next after the observed sequence.
In actual implementation, the model would be trained using backpropagation from estimated gradients e.g. via the parameter-shift rule \citep{schuld2019evaluating}. 
The model would produce for each shot a sample corresponding to the index of a single predicted word. 

Our proposed architecture is evaluated in a small-scale noiseless simulation. We consider a dataset of 7 sentences using a vocabulary of 11 unique words. We compare our proposal against two baseline models: one random uniform prediction model and one inspired by \cite{london2023peptide}. We compare performance between models trained on 5 sentences by evaluating on the remaining 2 sentences their perplexity \citep{jelinek1977perplexity}, for which a lower score indicates better performance.
A naive uniform random prediction on this dataset yields a perplexity of 11. Using a 9-qubit \cite{london2023peptide} model with 297 parameters, we achieve a perplexity of 8.15. Using a 9-qubit model that we propose with 172 parameters, we achieve a perplexity of 2.79.

To our knowledge, this is the first fully quantum sequential text generation architecture that is designed with the capabilities and limitations of current NISQ-era devices in mind. Our simulation results demonstrate the viability of the approach for implementation on actual hardware while achieving a reasonable level of perplexity.

\section{Attention in Quantum NLP Models}
\label{sec:attention}




So far we have discussed models for studying sentences as word / token sequences. Making such models scale
to longer sequences has always been a challenge: with $n$-gram models, the value of $n$ has always been small \cite[Ch 6]{manning1999foundations}; and RNN architectures including LSTMs, while accurate for short sequences, had 
trouble scaling to cover long-range dependencies \cite[Ch 15]{geron2019hands}.

\rev{\subsection{Attention in Classical LLMs}}

Attention is designed to address this problem.
The attention methodology was used to enhance an RNN sequence model for machine translation by
\cite{bahdanau2014neural}, which enabled the model to capture longer-range relationships as well.
 Although it still relied on the encoder-decoder paradigm, the bidirectional RNN architecture introduced in \cite{bahdanau2014neural} features a distinct context vector for each word in the sentence. Each context vector depends on a sequence of annotations which contains information about the entire sentence with a strong focus on the parts of the sentence surrounding the context vector's associated input word. The annotations are weighted according to an alignment model, which scores how well an output token matches inputs around a given position. 

\cite{vaswani2017attention} developed this approach further, 
demonstrating a system where transformer blocks incorporating attention, layer norm, multi-layer perceptrons, and residual connections,
fully replace recursive units. 
This Transformer model --- centered around \textit{scaled dot-product attention} --- made previous RNN-based encoder-decoder architectures obsolete when it demonstrated improved performance on various translation tasks.

Importantly, \cite{vaswani2017attention} adapted the Transformer architecture for use in text generation. Their model is auto-regressive, and at each step it consumes the previously generated symbols as additional input when producing new text. In addition, it is worthy to note that the Transformer was later adapted to the setting of computer vision, where it outperformed state-of-the-art convolutional neural networks in various image classification challenges \citep{dosovitskiy2020image}.

In general, ``an attention function can be described as mapping a query and a set of key-value pairs to an output, where the query, keys, values, and outputs are all vectors'' representing embedded tokens; the output is a weighted sum of the values, with the weights measuring the compatibility between corresponding query and key \citep{vaswani2017attention}. \textit{Self-attention} refers to computing attention coefficients intra-sequence, i.e., on the same input sequence. A key feature of self-attention layers is that they provide a mechanism for different tokens in the input sequence to interact, thereby allowing models to infer contextual information about individual tokens by weighing the importance of pairwise interactions; in other words, how much attention a given input token should pay to every other token in the sequence.

In particular, the ``Scaled Dot-Product'' attention layer featured in \cite{vaswani2017attention} computes the dot-products of the query with all the keys, normalizes according to the dimension of the query and key vectors, and then applies the softmax function to obtain the weights of all the pairs, which are the values. Rather then enumerate all the indices for summation, it is typical to write the 
lists of vectors as matrices, whereby the definition takes the common form
\[\mathrm{Attention}(Q, K, V)  = \mathrm{softmax}\left(\frac{QK^T}{\sqrt{d}}\right)V,\]
where $d$ is the embedding dimension, $Q$, $K$, and $V$ are matrices of size $wd$ where $w$ is the number of words / tokens in a sequence,
and the softmax function $x_i \rightarrow e^{x_i} / \sum_i (e^{x_i})$ is applied to each row. This formulates dot product attention as a matrix
multiplication ($O(w^2d)$), a softmax step ($O(w^2)$), and a final matrix multiplication ($O(wd^2)$).


A key advantage enjoyed by the Transformer over the previous RNN architectures is that this multiplication can be parallelized,
which computes the pairwise relationships between all the tokens in a sequence at once. In addition, the computational cost does not depend on the distance between tokens in the sequence, as in previous models. Together, these properties accounted for a drastic reduction in training time over sequential RNN models. The main drawback is that the computational complexity still scales quadratically in the number of tokens in a given sequence (roughly the number of words in a sentence). The problem of approximating or providing an alternative to self-attention with subquadratic complexity spawned its own burgeoning research field \citep{zaheer2020big,poli2023hyena}.

\rev{\subsection{Near-term quantum self-attention mechanisms}}

In hopes of improving this quadratic scaling, and since attention layers have become so successful as key components in state-of-the-art models for NLP tasks, various quantum approaches have been suggested. This section focuses on near-term quantum circuit designs.

\begin{figure}[!ht]
    \centering
    \includegraphics[width=0.43\textwidth]{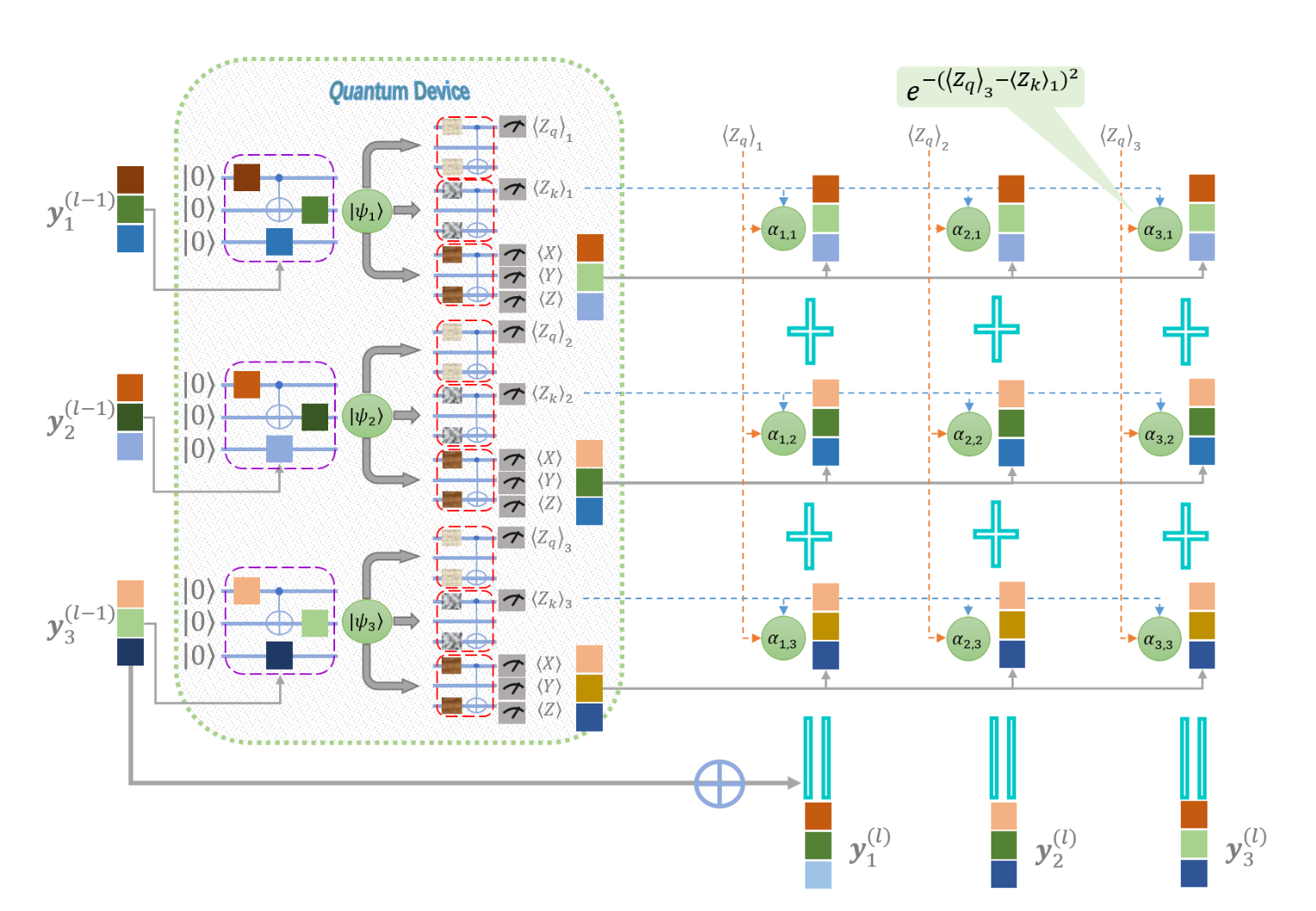}
    \caption{The Quantum Self-Attention Neural Network (QSANN) architecture proposed in \cite{li2022quantumselfattention}. The network features various consecutive self-attention layers. At the $(l-1)$st layer, the classical feature vectors $y_k^{(l-1)}$ are encoded into a high-dimensional qubit state space (circuits boxed in purple). The process is repeated three times. Then parametrized ansatze, with gate layout as in Figure \ref{fig:qsam_ansatz}, representing the query, key, and value transformations are applied (circuits boxed in red). The resulting states are measured and various expectation values are computed to produce the classical query, key, and value vectors. These are sent to a classical device for processing, where weights are computed using a Gaussian kernel and the results are averaged to obtain final attention coefficients.}
    \label{fig:qsann}
\end{figure}
\begin{figure}[!ht]
    \centering
    \includegraphics[width=0.43\textwidth]{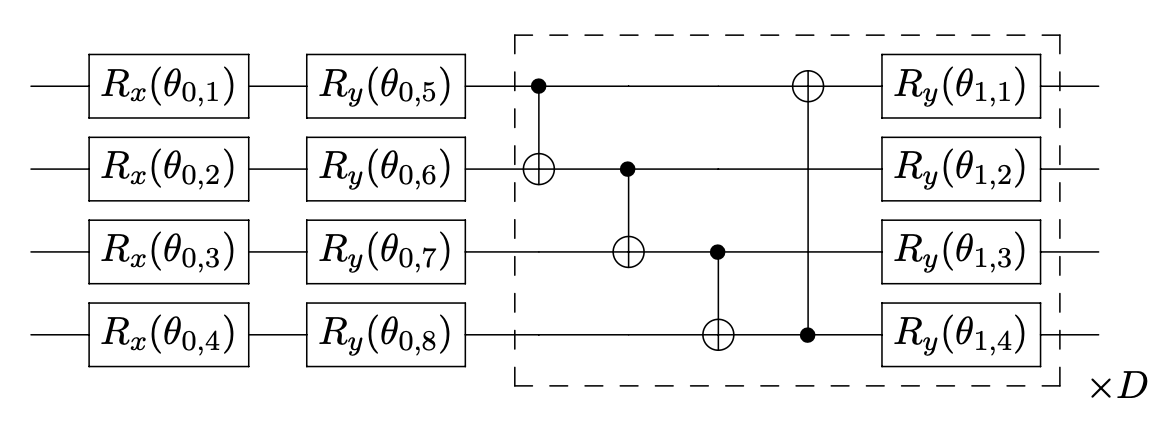}
    \caption{Parametrized ansatz implementing the query, key, and value transformations in \cite{li2022quantumselfattention}'s QSANN.}
    \label{fig:qsam_ansatz}
\end{figure}

\cite{li2022quantumselfattention} claim to offer the first quantum self-attention implementation, QSANN. By mapping encoded feature vectors into a high-dimensional Hilbert space using a quantum circuit, QSANN aims to extract correlations that are intractable classically. 
For an illustration see Figure \ref{fig:qsann}. First they construct an encoder circuit to load classical feature vectors onto an $n$-qubit quantum state; they use one classical feature vector for each token in the input sequence. The number of qubits $n$ is a hyper-parameter that should be adjusted as relevant to available hardware. Next they apply parametrized quantum circuits, with identical gate layouts but different parameter values in order to compute the query, key, and value vectors for each classical feature vector. The circuit layout is illustrated in Figure \ref{fig:qsam_ansatz}. 
At this stage the query, key, and value vectors are encoded as quantum states, so measurements must be made in order to extract useful information; the resulting query and key are the expectation values of the Pauli-$Z$ operators applied to the first qubit of the resulting states, and the value is a vector of expectation values of various Pauli operators. 
Attention scores are then computed on a classical device as a weighted average of the output values. Interestingly, \cite{li2022quantumselfattention} introduce a Gaussian kernel to compute the weights on the values vector; they claim the Gaussian kernel can more easily correlate quantum states with little overlap, which is needed, e.g., if two tokens are closely related in a sentence but their quantum state embeddings happen to be distant in the qubit state space.
The proposal of \cite{li2022quantumselfattention} still requires quadratic classical computation, and its main source of quantum advantage relies on using efficiently processing vectors in high-dimensional Hilbert space to unearth hidden relationships between embedded tokens.

\begin{figure}
    \centering
    \includegraphics[width=0.43\textwidth]{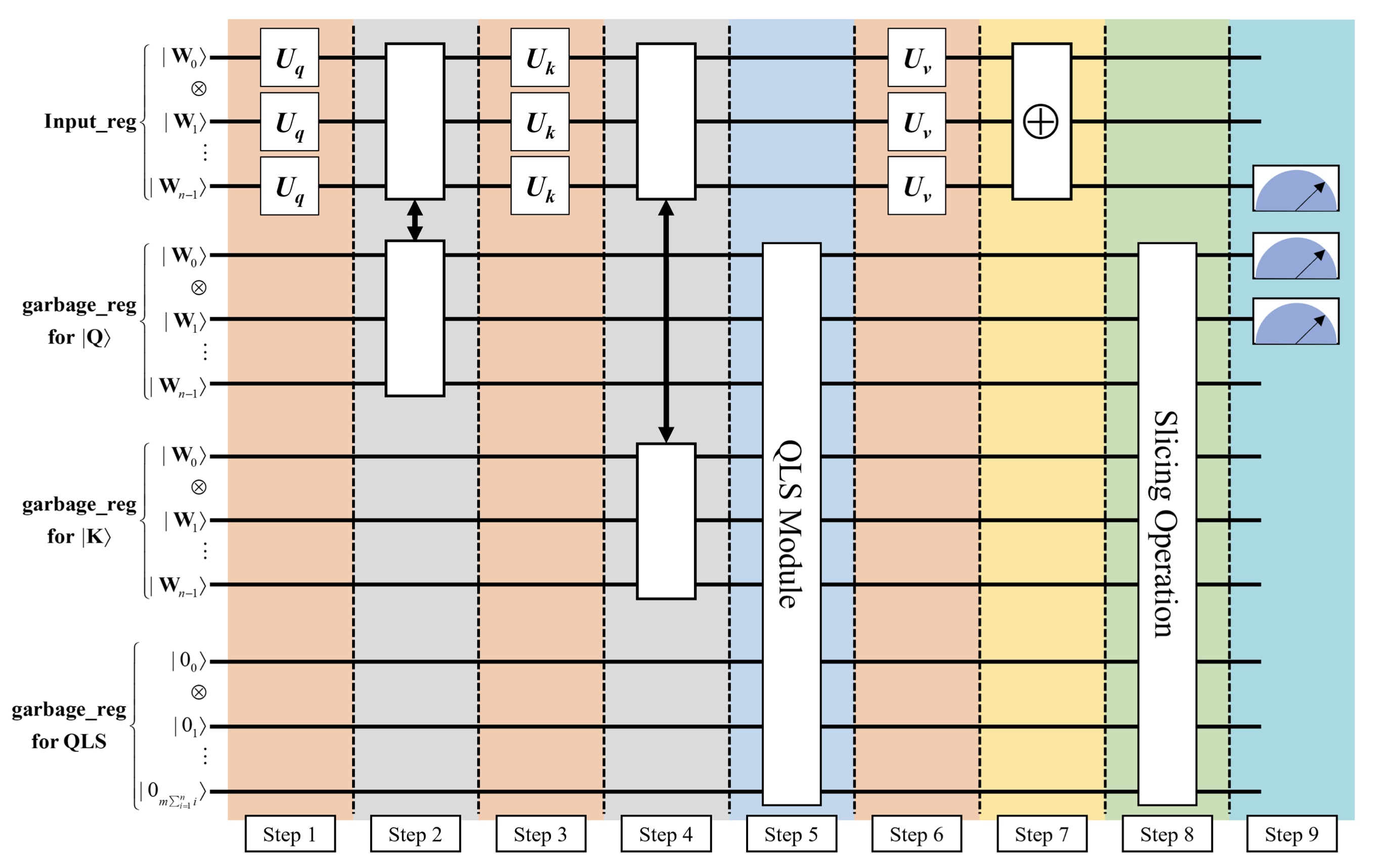}
    \caption{The Quantum Self-Attention Network (QSAN) introduced in \cite{zhao2022qsan}. This architecture uses ancilla qubits to hold intermediate results and proposes computing the attention coefficients entirely on the quantum processor, obviating the need for intermediate measurements. In addition, it features a slicing operation to reduce the number of measurements required.}
    \label{fig:qsan}
\end{figure}

\begin{figure}
    \centering
    \includegraphics[width=0.43\textwidth]{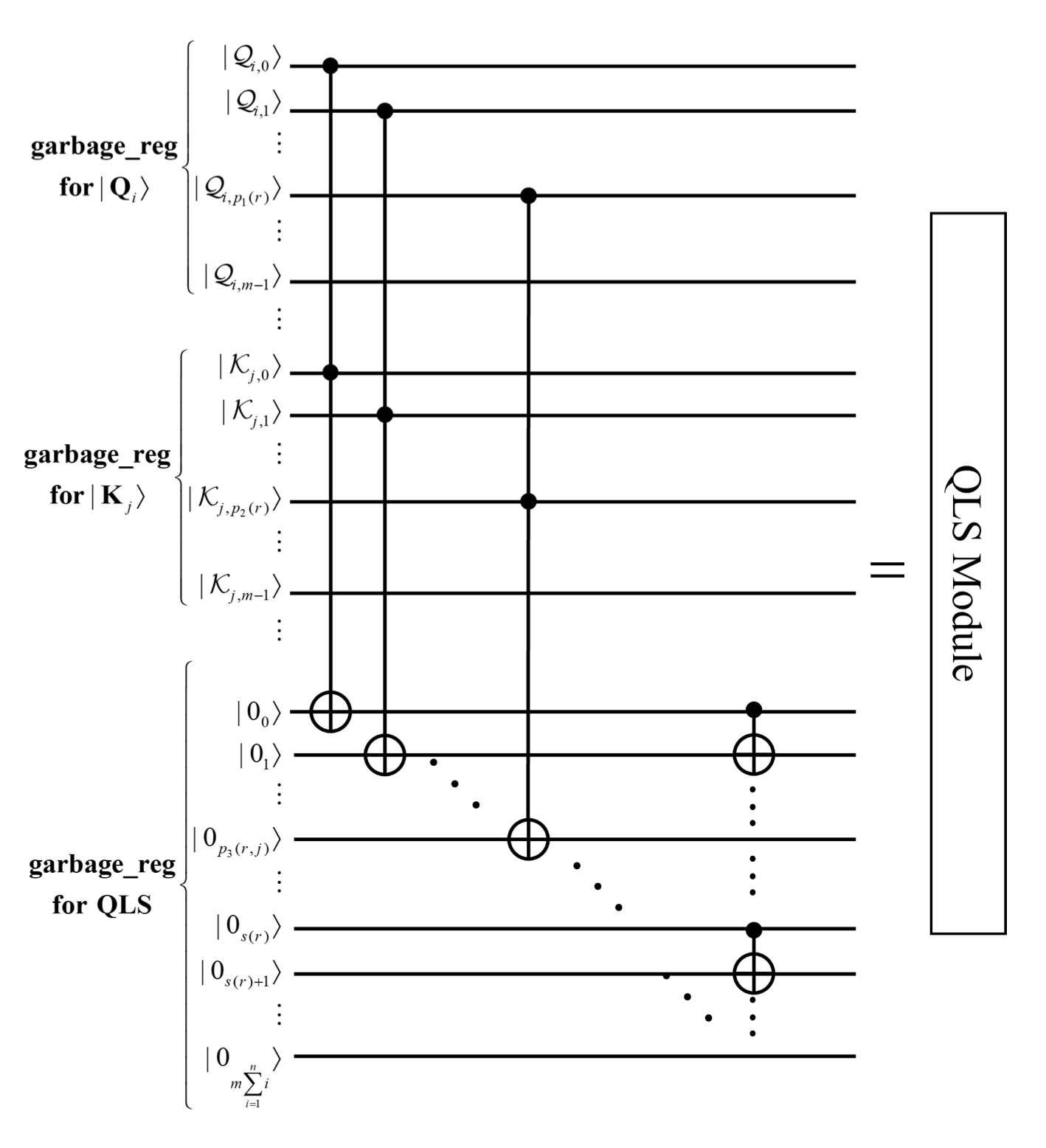}
    \caption{The Quantum Logic Similarity (QLS) module proposed in \cite{zhao2022qsan}, implemented as a sequence of Toffoli and CNOT gates.}
    \label{fig:qls_module}
\end{figure}

The work of \citet{zhao2022qsan} builds on these ideas, seeking quantum advantage in the same vein. By introducing various sets of ancilla qubits, the authors obviate the need to perform intermediate measurements during the attention computation. 
(This could be thought of as a more sophisticated example of the ancilla readout qubits pattern used in the QPOSTR
design of Section \ref{sec:qc-intro}.)
In this modality, query, key, and value quantum state-vectors are computed by applying parametrized ansatze and swapping onto ancilla registers sequentially, as shown in Figure \ref{fig:qsan}. Compatibility between query and keys is computed by a \textit{Quantum Logical Similarity (QLS)} module, which is implemented as a sequence of Toffoli and CNOT gates, as shown in Figure \ref{fig:qls_module}. This is a key step: it computes the overlap between query on keys directly on the quantum device, thereby improving on \cite{li2022quantumselfattention}. 

\begin{figure}
    \centering
    \includegraphics[width=0.45\textwidth]{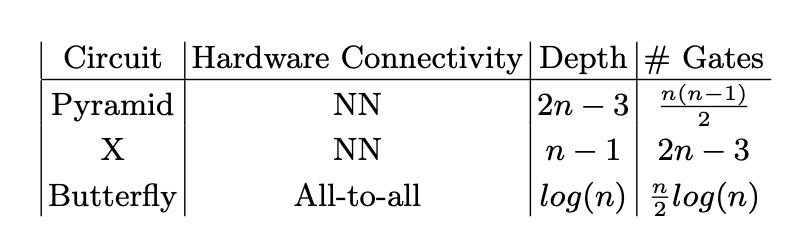}
    \caption{Computational complexity of the dot-product compatibility between query and keys using circuits with parallel two-qubit gates as proposed by \cite{cherrat2022quantumvision}'s quantum Vision Transformer.}
    \label{fig:qvit_attn_complexity}
\end{figure}

While these two proposals address quantum self-attention mechanisms in QNLP directly, \cite{cherrat2022quantumvision} proposes one for use in Vision Transformers for image classification, seeking quantum advantage in reducing the computational cost of the scaled dot-product attention calculation. Concretely, the authors introduce so-called \textit{orthogonal layers} to compute compatibility scores between query and keys on the quantum hardware; these layers efficiently implement parametrized transformations on encoded feature vectors, as described in Figure \ref{fig:qvit_attn_complexity}. The main novelty here is that \cite{cherrat2022quantumvision} use the unary encoding circuit to encode token feature vectors into the Hamming weight-1 subspace of the qubit state space. This encoding is advantageous because their orthogonal layers preserve the subspace, and they can be used to compute dot-products between query and keys in logarithmic time, assuming quantum gates can be applied in parallel. \rev{\citet{cherrat2022quantumvision} report preliminary results from simulation, and with a 6-qubit quantum processor.}

\rev{\subsection{Attention on Fault-Tolerant Quantum Computers}

The role of attention in systems like GPT \citep{brown2020language} has spurred more ambitious proposals,
including the recent preprints of \cite{liao2024gptquantum} and \cite{guo2024quantumtransformer}, which describe
large-scale versions of full transformer-inspired processes on fault-tolerant error-correcting quantum computers. \cite{guo2024quantumtransformer} use
quantum signal processing and singular value transformation techniques to apply a polynomial approximate softmax,
whereas \citep{liao2024gptquantum}, like \citep{zhao2022qsan}, replace softmax with a linear quantum alternative.

A main challenge for such proposals remains quantum I/O, which these papers acknowledge, but do not solve explicitly.
To extract all the weights from a matrix stored in a quantum state would typically require many more shots
than there are weights.
A smaller alternative is to output token indices directly. A natural NISQ-inspired suggestion is to adapt the
token index generation approaches from Section \ref{sec:seq} to transformer prototypes. While adding a token-generation head may be useful for circumventing I/O issues, the complexity and depth of the remaining bodies of the circuits in these proposals puts practical implementations beyond the NISQ era.

A quantum decoder, for
optimized search through a much larger space of long proposed token sequences, is proposed by \citet{bausch2021quantum},
which casts the problem as probabilistically branching tree-search, which is mathematically equivalent to probabilistic
grammar parsing. 
Even though the hardware capabilities for such quantum operations are still some years away, 
this connects optimizations in sequence generation from today's softmaxed probabilities, with traditional syntactic approaches 
to language modeling, which we discuss next.
}

\section{Syntactic Parsing and Logical Forms}
\label{sec:syntax}

The use of general AI techniques such as RNNs and attention has fueled much recent success with NLP,
partly because it has enabled much cross-fertilization between language and other kinds of data such as images, 
audio, and graphs. There are also more traditional NLP techniques, based on grammatical structures
found particularly in human language.

Natural languages (and artificial programming languages) express many structured relationships that
go beyond proximity in a sequence model. For example, in the sentence 
``Kim kicked the ball into the goal from half-way down the pitch, right past the goalkeeper, and scored'', 
the grammatical structure of English enables us to infer easily that the person who scored is Kim, in spite of there
being 16 words (including 4 other nouns) in between the noun {\it Kim} and the verb {\it scored}. 
Language grammar and syntax studies how these relationships are expressed and structured in different languages, and this field 
powerfully influenced much of theoretical and computational linguistics during the second-half of the $20^\mathrm{th}$ century,
particularly through the work of \citet{chomsky1957syntactic,chomsky1965aspects}.
In such a framework, syntax is the central generative system, on which other aspects of language like phonology and semantics
depend \citep[Ch 5]{jackendoff2002foundations}.

During the $21^\mathrm{st}$ century, the reliance of NLP techniques on grammatical rules
has declined. This is partly due to the success of statistical and machine-learned models that share methods with other
data-intensive areas of AI, and perhaps because the increasing preponderance of informal text created on smartphones 
immediately contradicts any assumption that the input to an NLP system should consist of distinct grammatical sentences.
In computational terms, best-performing parsing algorithms have included the CYK-parser which is worst-case $O(n^3)$,
while the more general attention mechanism discussed in the previous section has a baseline $O(n^2)$ performance, 
which robustly enables {\it more} resources to be targeted at important long-range relationships.
Hence in most large-scale NLP systems today, a grammatical parser is not an explicit component: and with the challenge
of reducing computational costs below quadratic, there would need to be a strong reason for requiring the
extra burden of a cubic computing step for any large-scale model.
When grammatical parsing {\it is} discussed, it is often as a historical challenge that LLMs can
solve quite effectively, not a contemporary challenge for building language models in the first place \citep{min2023recent}.

However, the use of grammatical parsers has been re-introduced in parts of {\it quantum} NLP,
motivated especially by a mathematical correspondence between the compositional rules of tensors
and categorial grammars demonstrated by \citet{coecke2010distributional}.
This framework is implemented in the \texttt{lambeq} system of \citet{kartsaklis2021lambeq} which is particularly 
designed for quantum computers \citep{lorenz2023qnlp}. In this system, a grammatical parser
prepares a parse-tree from a sentence, which is structurally mapped to a tensor
network which is compiled into a quantum circuit: so the quantum circuit encodes the
grammatical dependencies of the sentence, assuming its input data consists of uniquely-parsed sentences.

The parsing problem itself is also an interesting challenge for quantum computing,
for combinatoric reasons. Computationally, the problem can
be phrased as taking a list of words as input, and returning a parse-tree, which is a data structure saying
how the elements of the sentence are grouped together, and what grammatical role they play.
The two possible tree-structures for a 3-word sentence are shown in Figure \ref{fig:3_element_trees}.
For the simple case of a 3-word sentence, there are only 2 possible trees, one for the
structure $[[A\ B]\ C]$ and one for the structure $[A\ [B\ C]]$.
For a 4-word sentence, the five possible trees are are shown in Figure \ref{fig:4_element_trees}.
The number of possible parse trees for a sentence of length $n$ grows exponentially and is given by the {\it Catalan number}
$C_n = \frac{1}{n+1} \binom{2n}{n}$. 

\begin{figure}[t]
\begin{center}
\begin{tabular}{ccc}
 \Tree [.S [.NP [.D the ] [.N music ] ] [.VP [.V played ] ] ]  
 & &
 \Tree [.S [.NP [.NPlural people ]] [.VP [.V played ] [.NP [.N music ] ] ] ]
 \\ 
Left-branching tree & & Right-branching tree \\
\end{tabular}
\end{center}
\caption{The simplest nontrivial tree-parsing challenge is to distinguish between the two distinct
branching options for a tree with 3 leaf nodes. S = Sentence, N = Nouns, V = Verb, D = Determiner, P = Phrase.}
\label{fig:3_element_trees}
\end{figure}

\begin{figure}
    \centering
    \includegraphics[width=0.9\linewidth,trim=0 4cm 0 1cm,clip]{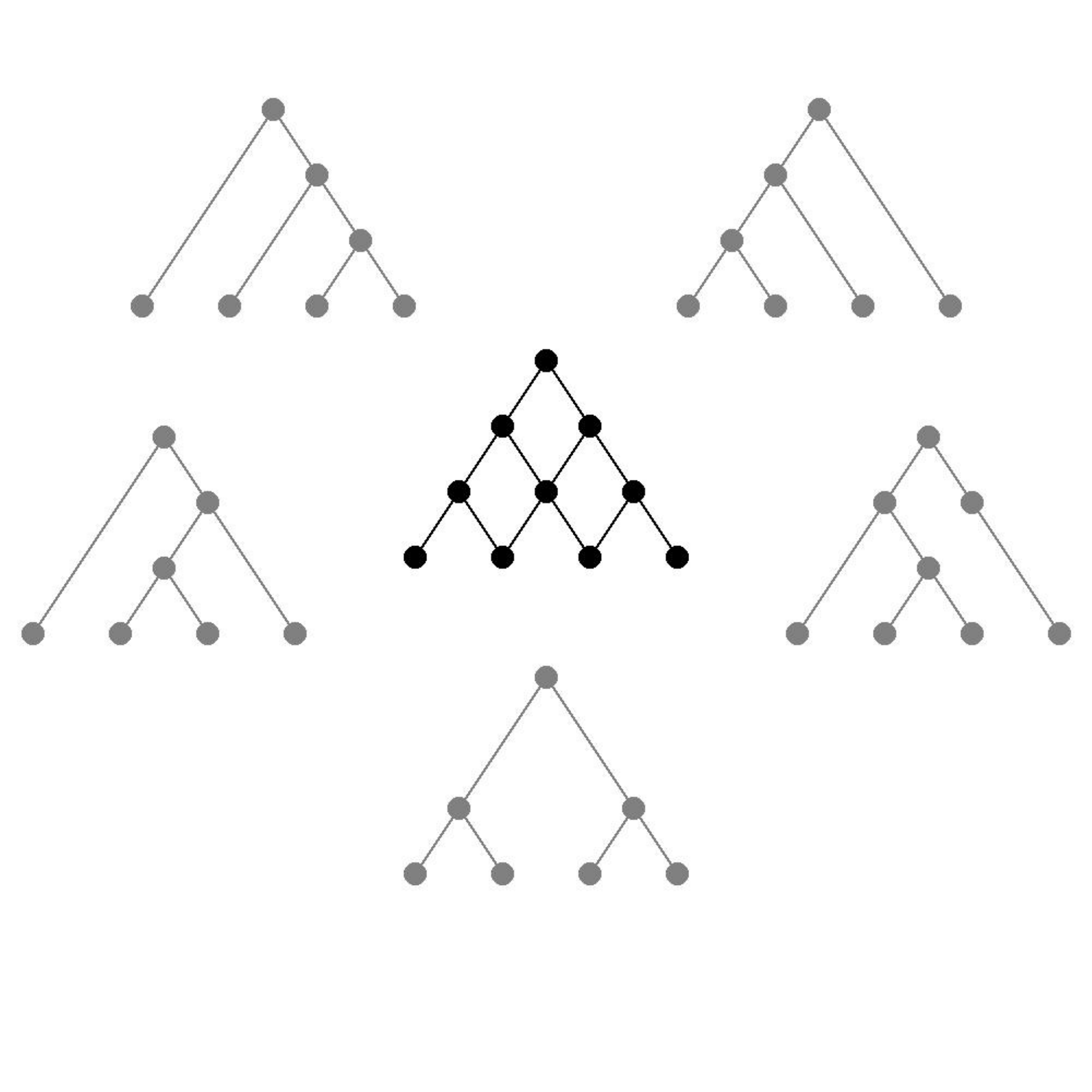}
    \caption{The five possible binary-branching trees over a 4-item sequence, each of which can be constructed
    by deleting unwanted branches and nodes from the connected lattice-graph in the center.}
    \label{fig:4_element_trees}
\end{figure}

Ideally, a parser will find an exact parse that accounts for the role of 
each constituent in the sentence, represented by a single tree-structure. 
Grammatical rules such as ``A sentence in English can be made of a noun phrase subject followed by a verb phrase''
can be represented as recipes like $\mathrm{S}\rightarrow(\mathrm{NP, VP})$, and a probabilistic phrase-structure grammar (PCFG)
may include many such rules, along with probabilities or weights learned from training data.
Various methods for parsing exist,
relying on dynamic programming and probabilistic techniques \cite[Ch 17]{jurafsky2023speech}. 
Multiple parses
are often possible, because (for example) prepositional phrases may attach in various locations.
The goal is to find the parse tree that maximizes the overall probability of the parse, computed from the 
combined probability of all the rules used in derivation.

The depiction in Figure \ref{fig:4_element_trees} emphasizes the parsing problem as a combinatoric challenge.
The search for a best parse looks like a kind of minimal spanning tree 
problem, which might be formulated as an Ising model, amenable to quantum optimization 
\citep{lucas2014ising}, if the weights or costs for visiting each node could be derived from the PCFG.

However, a key challenge with the parsing problem is that we start with labels only for the leaf nodes,
and until there is some hypothesis for the which internal nodes represent which syntactic chunks, there
are no estimates for the weights on internal links. (This is part of the motivation for the use of dynamic programming
in classical parsers.) Combinatorically, the problem is not exactly to find a minimum spanning tree, because
internal nodes that do not correspond to distinct grammatical phrases or {\it constituents} do not need to be visited.
(Moreover, the initial graph is directed: this precludes solutions where the leaf nodes are all visited 
by a zigzag path along the bottom, because upward steps are forbidden.)

Hence, quantum parsing does not appear to fit one of the known techniques of quantum combinatorial
optimization, but the problems are tantalizingly similar, and we hope this framing of the challenge
helps the language parsing problem to capture the interest of quantum researchers.

One potential benefit of quantum parsing is that a quantum system may represent more nuanced proposals than just 
returning a single best parse, or even a classical mixture (estimated probability distribution over different discrete parse proposals).
Rather than insisting that all language inputs must pass a parsing test before being processed, a syntactic parser that
exchanges quantum information with semantic components in parallel may become a different kind of asset in more parallel architectures.

In another interesting combination of quantum mathematics and natural language syntax,
it has been shown that tensor product networks
can encode grammatical structure more effectively than LSTMs for generating image captions \citep{huang2017tensor}.
Tensor product networks have also been used to construct an attention mechanism from which grammatical structure can be recovered
by unbinding role-filler tensor compositions \citep{huang2019attentive}.

\section{Facts and Language Generation}
\label{sec:facts}

Throughout its history, quantum theory has motivated new insights on probability
and randomness, how the potential and actual are related, and this has led to proposed models
for various aspects of human behavior and consciousness \citep{busemeyer2012quantummodels,atmanspacher2020consciousness}.
Rapid advancements in customer-facing AI systems, particularly conversational dialog systems
supported by large language models (LLMs), have raised many questions about how such
systems should be built and used, what should be expected of them, 
and about whether they exhibit conscious behavior. This section discusses some of the 
concerns and work in this area, including how quantum theory addresses the difference between
hypothetical and actual reality.

One of the key complaints about current LLMs is their propensity to \textit{hallucinate}, or produce sentences with factually false information that still correctly adhere to grammatical rules. While this behavior is in line with what generative models are statistically designed to do, in practice it goes against the public expectation of the AI agent as an all-knowing oracle, especially when manifested as an interactive, question-answering chat interface. Thus, several lines of research have emerged to address the usability issues caused by hallucination. In this section, we first review some of these research directions, and then take a step back to see how quantum computing relates to the philosophical issues that arise in understanding and using LLMs.

Two prominent realms of research are chain\hyp{of}\hyp \linebreak 
thought (CoT) prompting and retrieval augmented generation (RAG).
CoT, while often used in the context of complex reasoning tasks \citep{wei2022chain}, is believed to produce more self-consistent results and indeed can be improved by explicitly encouraging self-consistency \citep{wang2023self-consistency}. However, in the setting of factuality, errors early in the chain may yield incorrect results even with consistent reasoning later in the chain, and the explanation of reasoning provided by the chain may not be correct \citep{turpin2023language}. 

Factual grounding via RAG may avoid some issues still present in CoT, often at the cost of either additional training or larger number of inferences. RAG can be largely classed into two paradigms, \textit{a priori} and \textit{post-hoc}, though there is not reason that these techniques could not be combined or even performed iteratively in a loop.
\cite{lewis2020retrieval-augmented, guu2020retrieval} explore \textit{a priori} RAG using learned embeddings stored in a database and accessed with a learned neural retriever. \cite{borgeaud2022improving} scale up the database and shrink the language model, matching SotA performance and illustrating that the world knowledge of a LLM is more separable from its linguistic ability than previously demonstrated. Together, this line of work suggests separability and knowledge base scaling as one possible path forward for utilizing word embeddings to reduce hallucination. While these \textit{a priori} methods show promise, they modify the generation pipeline and require additional training.

Post-hoc text editing methods \citep{thorne2021evidence-based, balachandran2022correcting, schick2022peer} are seeing interest for use with LLMs in part due to the resources required to even fine-tune modern LLMs, let alone pre-train them from scratch. For example, \cite{gao2023rarr} uses the abilities of pre-trained language models and existing information retrieval systems to edit and verify generated text, but requires running several rounds of inference on the base LLM. This trades off a zero fixed cost of using a pre-trained model for a higher variable cost of inference.

In such paradigms, one natural place where quantum computing has potential to enhance performance is in accessing and designing the knowledge base. Grover's algorithm \citep{grover1996fast} yields a quadratic speedup over classical methods for searching a database, and hence forms the backbone of quantum information retrieval systems \citep{giri2017review}. Going one step further, \cite{pronin2023synthesis} propose using Grover's search to inform the design of the database itself, and specifically target vector databases as the intended application. However, they do not directly compare the quadratic speedup with the complexities achievable by using efficient classical data structures, and a potential interesting direction in this area is the question of how (if possible) to design a practical hybrid knowledge base that combines the best of both classical and quantum processing.

\begin{figure}
    \centering
    \includegraphics[width=\linewidth]{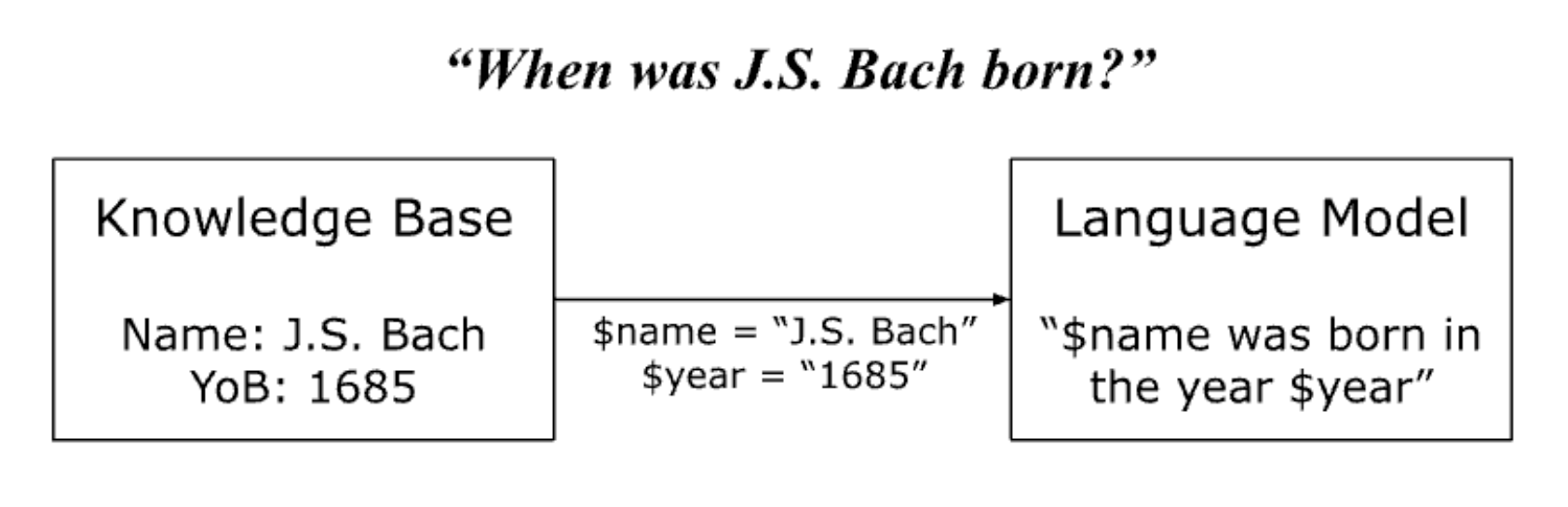}
    \caption{A traditional division of responsibilities between a knowledge base and a language
    model that cooperate in generating the answer ``J.S. Bach was born in 1685.''}
    \label{fig:kb_lm}
\end{figure}

Although techniques such as retrieval-based and knowledge-based generation are a new 
area in the present-day context of fixing LLMs, methods such as that of \cite{borgeaud2022improving}
hearken back to an older class of designs where the facts are 
stored in a knowledge base,
and the language model is effectively a source of templates, not of facts. 
The spotlight on language generation in the past few years has refocused work on such methods,
and how best to combine them with LLMs \cite{zhang2023controllable}.
For example, in Figure \ref{fig:kb_lm}, a knowledge base is used to find the variable that satisfies the question ``When was J.S. Bach born?'' (the answer being ``1685''), and then a language model is used to express 
this as the sentence ``J.S. Bach was born in the year 1685.'' (A standard early use of such designs 
was in mail merging, where a template for a message is combined with a list of different names to generate
personalized messages.) Such a language model can just as easily generate the sentence 
``J.S. Bach was born in 1985'', not because it's hallucinating, but because it's working correctly with a different knowledge base. 

More generally, probabilistic language models are designed to note that {\it Wednesday} and {\it Thursday} are similar, and so having seen the phrase ``Let's meet on Wednesday'', 
the model should judge the phrase ``Let's meet on Thursday'' to be similarly plausible. 
Saying that such a probabilistic model ``hallucinates'' when it generates an untrue sentence
reflects a fundamental misunderstanding of probabilistic models. Sampling from a probabilistic model
is like rolling dice: if we previously observe a 3 and the dice-roll gives a 4, the dice aren't 
hallucinating a 4 instead of the ``true'' value of 3. The problem lies in the assumption 
that plausible probabilistic samples of {\it language} should correspond to {\it facts} at all.

The assumption that language expertise and factual reliability {\it should} go together is easy to make, especially since a significant amount of actual knowledge is conveyed both explicitly and implicitly through writing.
In the phrase ``Dave beat John'', we might ask ``Which Dave and which John?'' before assessing its
truthfulness: but sometimes words take on fixed unique meanings in particular situations, so that
if someone says ``Caesar beat Pompey'', we automatically assume they mean two particular people 
from the 1st century BC, and if they said ``Pompey beat Caesar'', that would be considered untrue.
However, speaking strictly in the sense of language modelling, a model that also generates ``Pompey beat Caesar'' sometimes as well as ``Caesar beat Pompey'', is arguably {\it better}, because it generates
a more comprehensive variety of perfectly fluent and plausible sentences.

The practice of generating text from {\it just} a language model was popularized by successful machine
translation systems \citep{bahdanau2014neural}. With machine translation, it makes sense that the system
is {\it not} responsible for factual accuracy, because this is the user's responsibility. In concrete terms,
a correct translation of ``J.S. Bach was born in 1985'' from English to German might be ``J.S. Bach wurde 1985 geboren'', not ``Input error: prompt contains factual inaccuracy.'' Gradually models such 
as GPT demonstrated that a whole range of prompts, not just translation targets, 
could elicit plausible and fluent responses \citep{brown2020language}. The Chomskian program claimed grammatical fluency 
as the heart of language decades ago: today, we are seeing that this fluency is 
another aspect of human behavior that computers can
mimic effectively; and the ability to assemble erudite text has become one of the most impressively-solved parts of AI, 
sometimes leading to problems elsewhere.

Quantum theory intersects with these topics even more fundamentally, by explicitly distinguishing the possible from the actual.
A quantum circuit has many possible outcomes that could be observed, but only one outcome is observed 
when measured: and this fixes the hypothetical situation so that the same outcome is observed next time.
A multiplicity of possibilities can become a single fixed event \citep{heisenberg1958physics}.
Formal similarities between
this process and language ambiguity were noted by \citet{widdows2003contextmodel}, and the 
quantum economic theory of \cite{orrell2020quantumeconomics} is based on the use of quantum information
to model beliefs about values, and classical information to model amounts of money agreed in fixed transactions. 

The problem of distinguishing things that {\it might} happen from things that {\it do} happen was behind
some of the controversies of early classical mechanics. Leibniz discussed the notion of possible worlds,
and maintained that there must be a rational necessity behind (God's) choosing this world \citep{rescher1996leibniz}. 
Newton's belief in absolute space implied a fixed zero-point or origin, and
Leibniz argued that this implied that God must have made an arbitrary choice without a necessary reason,
which was unacceptable \citep{bouquiaux2008leibniz}. Such considerations of necessity vs. contingency and their
relationship to past, present, and future in time, go back at least to the
famous sea-fight discussion in Aristotle's {\it De Interpretatione} \citep{mckeon1941aristotle}. 

The notion that there are different possible worlds where a macroscopic event did or did not happen,
that one of those worlds is chosen based on a small local decision, and this possible world thus
{\it becomes} the actual world, was thrust into the limelight by quantum mechanics.
The implication of superposition and
large-scale randomness was troubling to Einstein (``God does not play dice!'')
and Schr\"{o}dinger, whose famous paradoxical cat was designed to illustrate the absurdity of
quantum mechanics in large-scale reality, where ``the working of an organism requires exact physical laws'' \cite{schrodinger1944life}.
By contrast, Bohr and Heisenberg supported the Copenhagen Interpretation, where
the wave-function represents real possibilities, and 
{\it ``the transition from the “possible” to the “actual” takes place during the act of observation" \citep[Ch 3]{heisenberg1958physics}.}

Some of the challenges inherent in large stochastic problems,
like weather forecasting, are thus philosophically 
related to key questions of how one possible future is selected and becomes the past.
Quantum mechanics does not completely answer this question, but it does better than classical mechanics,
where the assumption of a deterministic universe avoids the problem.
Heisenberg's analysis of where different uncertainties come from, and how we should think 
about them, has useful insights including 
{\it ``This probability function represents a mixture of two things, partly a fact and partly our knowledge of a fact" 
\citep[Ch 3]{heisenberg1958physics}.} This does not tell us how to fix language models, but  
it is a good reminder that our ways of stating and communicating facts are entirely human.
Practically, it helps to understand probabilistic language models as generators of hypothetical utterances,
rather than factual statements, and the generative nature of language models is precisely what enables them to go smoothly from data they encountered to data they might just as well have encountered.
In a sense, a large language generator is a kind of hypothesis-generator with 
the gift of the gab. Language models do this task very well, but this should never have convinced us
that a model will generate truthful language without an independent source of knowledge. 
With these considerations, quantum theory has some insight on potential solutions to 
improve language modeling systems, and at least guards against mistakes that arise from
over-deploying hypothesis-generation systems without suitable observation processes.

\section{Conclusion}

We have taken a whirlwind tour of the state of quantum NLP, seeing the potential and limitations of using quantum computers for understanding language. While we recognize this overview, like any other, cannot be fully comprehensive, we hope that it is nonetheless useful for both the theoretician and practitioner alike.

We reviewed fundamentals of gate-based quantum computing, and from here moved into understanding how these low-level structures and concepts can be used to efficiently encode basic units of language, i.e. text. From there, we built into progressively higher-level concepts, roughly following the hierarchy found in classical NLP.

Through this journey, we have seen how the current scale of applications for quantum NLP on actual hardware has not yet matched that of classical computing techniques. However, quantum methods being developed at the small scale show promise for use on intermediate scale problems as hardware continues progressing, and quantum models that have been shown to be more expressive than their analogous classical counterparts hold potential at large scales.

In the meantime, methods from quantum theory continue to inform AI. During the 2010s, vectors and tensors became
a common mathematical toolset permeating AI, and the adaptation of tensor network methods for scalability continues this
theme.
We have especially focused on the topical problems that current classical LLMs face. Here, quantum theory has much philosophical guidance to offer on the issues of assessing factuality and sequential inference.

\small
\bibliographystyle{spbasic}      
\bibliography{ionq}   

\end{document}